\documentclass[12pt,epsfig,color]{article}
\usepackage{amsmath}
\usepackage{epsfig}
\usepackage{amssymb}
\input{epsf}
\def\beq{\begin{equation}}
\def\eeq{\end{equation}}
\def\beqa{\begin{eqnarray}}
\def\eeqa{\end{eqnarray}}

\newcommand\as{\alpha_{\mathrm{S}}}
\newcommand\f[2]{\frac{#1}{#2}}
\def\la{\lambda}

\def\beq{\begin{equation}}
\def\eeq{\end{equation}}
\def\beeq{\begin{eqnarray}}
\def\eeeq{\end{eqnarray}}
\def\to{\rightarrow}
\def\nn{\nonumber}

\def\b0{b_0}
\def\bone{b_1}

\def\th{\hat{\tau}}

\def\qqb{q \bar{q}}
\def\ttb{t \bar{t}}

\begin{document}

\begin{titlepage}
\renewcommand{\thefootnote}{\fnsymbol{footnote}}
\begin{flushright}
BNL-NT-08/13 \\
YITP-SB-08-15
     \end{flushright}
\par \vspace{10mm}
\begin{center}
{\large \bf
Threshold Resummation for the Top Quark Charge Asymmetry}

\end{center}
\par \vspace{2mm}
\begin{center}
{\bf Leandro G. Almeida${}^{\,a}$,}
\hskip .2cm
{\bf George Sterman${}^{\,a}$,}
\hskip .2cm
{\bf Werner Vogelsang${}^{\,b}$  }\\[5mm]
\vspace{5mm}
${}^{a}\,$C.N.\ Yang Institute for Theoretical Physics,
Stony Brook University, \\
Stony Brook, New York 11794 -- 3840, U.S.A.\\[2mm]
${}^{b}\,$Physics Department, Brookhaven National Laboratory, 
Upton, NY 11973, U.S.A.\\

\end{center}


\par \vspace{9mm}
\begin{center} {\large \bf Abstract} \end{center}
We study the QCD charge asymmetry in $t\bar{t}$ production at
the Tevatron. We investigate the role of higher orders in perturbation
theory by considering the resummation of potentially large logarithmic 
corrections
that arise near partonic threshold.   This requires us to employ the
rapidity-dependent anomalous dimension matrices that describe color mixing
due to soft gluon emission in both quark- and gluon-initiated processes.
The charge asymmetry appears directly in the resummed cross section at 
next-to-leading logarithm (NLL), and we find that the first-order 
expansion of the NLL resummed charge asymmetry reproduces
the known fixed-order result for the asymmetry well. 
Beyond its lowest order, the asymmetric component of the cross 
section is enhanced
by the same leading-logarithmic threshold corrections as the total cross
section.   As a result, the charge asymmetry is  robust with respect to 
the higher-order
perturbative corrections generated by threshold resummation.
We observe that the asymmetry increases with pair mass and with
scattering angle.

\end{titlepage}

\setcounter{footnote}{1}
\renewcommand{\thefootnote}{\fnsymbol{footnote}}


\section{Introduction}

Heavy quark pair production in hadronic collisions is important both for 
accurate tests of the Standard Model and in searches for New Physics. 
Thanks to the large scale set by the mass of the heavy quark, one can often 
use QCD perturbation theory
in obtaining predictions for heavy-flavor production,
employing the factorization of the hadronic cross section into parton
distribution functions and perturbative short-distance cross sections. 

Top quark production may provide an arena for testing possible extensions 
of the Standard Model \cite{Lane:1995xx,Brooijmans:2008se,soni,bernreuther}. 
One particularly interesting observable in 
$t\bar{t}$ production that has been considered in this 
context~\cite{tree,Antunano:2007da} is the {\it charge asymmetry}
(or, forward-backward asymmetry), which is 
obtained by comparing the rate for producing a top
quark at a given angle to that for producing an anti-top at the 
same angle. Electro-weak processes, as well as processes in many
extensions of the Standard Model, may produce a charge asymmetry
at Born level. QCD, on the other hand, being a purely vector theory, 
does not produce a charge asymmetry in the lowest-order (LO) processes
$q\bar{q}\to t\bar{t}$ and $gg\to t\bar{t}$. The charge asymmetry
thus has the potential of probing or constraining possible tree-level
axial couplings of the gluon
\cite{tree,Antunano:2007da} at the Tevatron or the LHC.

Starting at order $ \as^3$, however, QCD itself contributes to the charge 
asymmetry, through $\qqb$ annihilation $ \qqb \to \ttb(g)$  and flavor 
excitation, $q g \to q \, \ttb$~\cite{Halzen:1987xd}. This happens 
through diagrams in which two separate fermion lines (one of them 
the top quark line) are connected by three gluons. This phenomenon,
which is also well-known in QED~\cite{Brown:1973ji}, enters with 
the combination $d_{abc}d^{abc}$ of the symmetric QCD structure constants.
The same mechanism for light flavors has been found to generate a
strange quark $s-\bar{s}$ asymmetry in the nucleon sea~\cite{Catani:2004nc}.
The charge asymmetric part is contained in the full next-to-leading order
(NLO, ${\cal O}(\as^3)$) calculations of the differential
heavy-flavor production cross 
section~\cite{Beenakker:1988bq,Nason,Marco}. Since this is the order
at which the effect arises for the first time, we will usually refer
to the charge asymmetric part arising at ${\cal O}(\as^3)$ as LO.
There have been detailed calculations and phenomenological studies of 
the QCD top quark charge asymmetry at the Tevatron (or the LHC), both 
for the inclusive case, $p\bar{p}\to \ttb X$~\cite{Antunano:2007da,Kuhn}, 
and for associated-jet final states,
$p\bar{p}\to \ttb \, {\mathrm{jet}}X$~\cite{Bowen,Dittmaier}.
In particular, Ref.~\cite{Dittmaier} provides the full NLO 
(${\cal O}(\as^4)$) corrections to $p\bar{p}\to \ttb \, {\mathrm{jet}}X$.

Very recently, first measurements of top quark charge asymmetries 
have been reported by the Tevatron collaborations~\cite{D0,CDF}. In the 
inclusive case, asymmetries of $12 \pm 8 \, {\mathrm{(stat.)}} \pm
{\mathrm{1(syst.)}}\%$ and $23 \pm 12 \, {\mathrm{(stat.)}} \pm
{\mathrm{6(syst.)}}\%$ were found by D0~\cite{D0} and CDF~\cite{CDF}, 
respectively. Even though experimental uncertainties are evidently
still large, this is a very encouraging first step that motivates further 
theoretical investigations. In the present paper we improve the theoretical 
framework for the case of the inclusive charge asymmetry by examining
the effects of QCD threshold resummation. This will provide insight
into the important question of how robust the asymmetry is with respect
to higher order QCD corrections. Our study requires us to 
implement NLL resummation in heavy quark production at fixed angle and 
rapidity~\cite{KS,Kidonakis:1998ur}. Earlier phenomenological studies of the 
threshold-resummed $t \bar{t}$ cross section, 
which however did not focus on the 
charge asymmetry, may be found in~\cite{topresu,BCMN,KLMV,Matteo,MU}.

As is well-known, when the initial partons have just enough energy to 
produce a $t \bar{t}$ pair, the phase space available for gluon 
bremsstrahlung nearly vanishes, giving rise to large logarithmic 
corrections to the partonic cross section. For example, if we consider 
the cross section for $\ttb$ production at fixed pair invariant mass,
this partonic threshold is reached when the pair invariant mass equals the 
partonic center of mass (c.m.) energy, $M_{\ttb}^2=s$. At the $n$th order
of perturbation theory, the large threshold corrections arise as 
$\as^{2+n}\left[ \log^m(1- \th)/(1-\th)\right]_+$ with 
$m \leq 2 n -1$, where $\th=M_{\ttb}^2/s$ and the ``+''-distribution
will be reviewed below. The maximum value,
$m=2n-1$ corresponds to the leading logarithms
(LL), $m=2n-2$ to next-to-leading logarithms (NLL), and so forth.
Near threshold, the perturbative calculation produces potentially large 
corrections at all orders in the strong coupling, $\as$. These corrections 
are addressed by threshold resummation. This
is particularly relevant for the Tevatron case, where the hadronic
c.m. energy is not too much larger than twice the
top mass, $2m_t$, so that $\th$ is on average rather close to unity. 
Related considerations also apply at the LHC when the pair is
produced with $M_{t\bar{t}}^2\gg4m_t^2$.

For heavy quark production, threshold resummation has been derived to 
NLL accuracy~\cite{BCMN,KS}, and recently to next-to-next-to-leading 
logarithm~\cite{MU}. The results of~\cite{KS} have been presented
for arbitrary c.m. scattering angle of the produced
top quark, which makes it possible to obtain a resummed charge asymmetry
from them. Among the processes that contribute to the charge asymmetry, 
only $\qqb$ annihilation contains threshold logarithms, while the 
flavor excitation $qg$ process is suppressed near threshold. As we
shall discuss in some detail, it turns out that the leading logarithms 
in the charge asymmetric part of  $\qqb$ annihilation cancel
at ${\cal O}(\alpha_s^3)$. This is 
because the charge asymmetric part is a difference of cross sections 
with the top or the anti-top produced at a certain angle, and the leading 
logarithms enter in association with the ${\cal O}(\as^2)$ $\qqb \to \ttb$ 
Born process, which is charge symmetric. 
We shall return to this point below, and will find that 
beyond ${\cal O}(\as^2)$  the charge asymmetric
cross section is enhanced by the same threshold logarithms as the
symmetric one. We also note that the $gg$ fusion process 
is charge symmetric to all orders; nonetheless its resummation can be 
relevant also for the charge asymmetry as it contributes to the 
denominator of the asymmetry and may thus dilute it somewhat. 
This effect does not lead to significant suppression, however,
because of the higher-order threshold enhancements to the asymmetric 
cross section, which we will exhibit below.

The remainder of this paper is organized as follows. In Sec.~2 we give 
the basic formulas associated with the charge asymmetry and discuss
the near-threshold behavior at ${\cal O}(\as^2)$. In Sec.~3 we present
the relevant expressions for the NLL resummed $t\bar{t}$ cross section
as a function of the $t\bar{t}$ pair invariant mass and the top c.m. 
scattering angle. Section~4 presents our phenomenological results for 
Tevatron kinematics, and we summarize our findings in Sec.~5.

\section{Perturbative Cross section, Charge Asymmetry, and 
Partonic Threshold}

We consider inclusive $t\bar{t}$ production in hadronic collisions, 
\beq
H_A( P_A) + H_B (P_B) \to t (p_t) \bar{t}  (p_{\bar{t}})+ X(p_X),
\eeq
where we have indicated the momenta. We introduce the invariant mass 
squared of the $t\bar{t}$ pair: $M^2_{t \bar{t}}=(p_t+p_{\bar{t}})^2$, and
the variable $\tau\equiv M_{t \bar{t}}^2/S$ 
with $S=(P_A+P_B)^2$. The factorized 
cross-section for the process is written in terms of convolutions of
parton distributions $f_a^{H_A}$ and $f_b^{H_B}$ for partons $a$,$b$
in hadrons $H_A,H_B$, respectively, with perturbative partonic 
hard-scattering cross-sections $\hat{\sigma}_{ab}$:
\beeq \label{eq2}
\frac{d^2 \sigma^{H_A H_B \to t \bar{t} X}}
{dM^2_{t \bar{t}}  \, d \cos \theta}&=&
\sum_{ab} \int_0^1 dx_a \int_0^1 dx_b \, f_a^{H_A}
(x_a,\mu^2) f_b^{H_B}(x_b,\mu^2) \, 
\frac{1}{s}
\frac{d^2 \hat{\sigma}_{ab}(\th,\cos \theta,\mu^2/s)}
{d \hat{\tau} \,d \cos \theta},
\eeeq
where the sum runs over all partonic subprocesses that produce top quark 
pairs. We have introduced the partonic variable $\th =M^2_{t \bar{t}}/s=
\tau/ x_a x_b$. We define $\theta$ as the production angle of the top 
quark in the partonic center-of-mass frame. We note from the outset
that this is not the definition adopted in the Tevatron experiments,
where the asymmetry is considered as a function of the rapidity
difference $\Delta y_t$ of the $t$ and $\bar{t}$. However, for LO 
kinematics, the two definitions are directly related through~\cite{CDF}  
$\tanh(\Delta y/2)=\beta \cos\theta$, with $\beta=\sqrt{1-4 m_t^2/s}$
the top quark velocity. As the partonic threshold regime is characterized
by LO kinematics, we expect our resummed results below 
to be very faithful representations also 
of the effects expected for the Tevatron definition 
(see also~\cite{Antunano:2007da}). In fact, we
have found that for quantities integrated over angle the charge asymmetries
for the two definitions agree at the level of about $2\%$ or better.
The integration limits in Eq.~(\ref{eq2}) are determined by
the conditions~\cite{KLMV} $s\geq M^2_{t \bar{t}}
\geq 4 m_t^2/(1-\beta^2 \cos^2 \theta)$. Finally, the scale $\mu$ denotes the 
factorization and renormalization scales, which we take to be equal 
throughout this study. 

We next define the charge-asymmetric and charge-averaged cross sections:
\beeq
\frac{d \Delta \sigma}{dM_{t\bar{t}}^2 d\cos \theta} &\equiv& 
\frac{1}{2} \left\{
  \frac{d^2 \sigma^{H_A H_B \to t \bar{t} X}}{dM_{t\bar{t}}^2 d\cos \theta}-
 \;  \frac{d^2 \sigma^{H_A H_B \to \bar{t} tX}}{dM_{t\bar{t}}^2 d\cos \theta} 
\right\},  \nonumber \\
 \frac{d \bar{\sigma}}{dM_{t\bar{t}}^2d\cos \theta}&\equiv& 
\frac{1}{2} \left\{  
 \frac{d^2 \sigma^{H_A H_B \to t \bar{t} X}}{dM_{t\bar{t}}^2 d\cos \theta} +
\frac{d^2 \sigma^{H_A H_B \to \bar{t} t X}}{dM_{t\bar{t}}^2 d\cos \theta}  
\right\} , \; \nonumber
\eeeq
and the corresponding charge asymmetry
\beeq
A_c(M_{t\bar{t}}^2,\cos \theta) &\equiv&
\frac{ d \Delta \sigma}{d \sigma} \; .
\eeeq

To lowest order (LO), $t\bar{t}$ pairs are produced by the processes
$q\bar{q}\to t\bar{t}$ and $gg\to t\bar{t}$. These produce the top and
the anti-top evenly at a given production angle $\theta$, so that
the charge asymmetry vanishes. Beyond LO, however, $\qqb$ annihilation
as well as the flavor excitation process $qg\to t\bar{t}q$ have
charge asymmetric contributions~\cite{Kuhn}, while $gg$ scattering 
remains symmetric. For the $\qqb$ annihilation process, the asymmetry 
arises from three gluons connecting the light-quark and
the top-quark lines. In the case of inclusive $\ttb$ production, 
the asymmetry receives contributions from real diagrams for
$\qqb\to \ttb g$ and from virtual corrections in $\qqb\to \ttb$.
The charge asymmetric pieces for the ${\cal O}(\as^3)$
subprocesses have been computed in detail in~\cite{Kuhn}; they are
also included of course in the full next-to-leading order (NLO) 
calculations of the top quark cross 
section~\cite{Beenakker:1988bq,Nason,Marco}. 

As we discussed in the Introduction, large double- and single-logarithmic 
corrections arise at higher orders in the partonic cross sections for 
$q\bar{q}$ annihilation and $gg$ fusion when $\hat{\tau}$ becomes large, 
that is, when $M_{t\bar{t}}^2\sim s$. The structure of the NLO
terms in the $\qqb$ annihilation cross section becomes, for example,
\beeq \label{plus}
\frac{d^2 \hat{\sigma}^{{\mathrm{NLO}}}_{\qqb}(\th,\cos \theta)}
{d \hat{\tau} \,d \cos \theta} &=& 
C_1 ( \theta) \delta (1- \th) + C_2(\theta) 
\left( \frac{1}{1-\th}\right)_+ + C_3(\theta) 
\left( \frac{\log (1-\th)}{1-\th} \right)_+ + \ldots \; ,
\eeeq
where the ellipses denote terms that are finite near partonic threshold.
We have suppressed the dependence on the factorization/renormalization
scale, as we will often do in the following. The ``+''-distribution 
is defined as usual by 
\begin{equation}
\int_x^1 dz \,\left[g(z)\right]_+ \, f(z) =
\int_x^1 dz \,g(z) \,\left(f(z)-f(1) \right) - f(1)\, \int_0^x dz \, g(z)\; .
\end{equation}
The coefficients $C_i$ may be found in Ref.~\cite{KLMV}. 
It turns out that only 
$C_1$ and $C_2$ possess charge-asymmetric pieces, while the double-logarithmic
part associated with $C_3$ is symmetric and thus cancels in the 
asymmetry at lowest order.
This is a result of the factorization of collinear logarithms,
which do not interfere with color flow in the hard scattering
\cite{KS}.    
As a result, the charge asymmetric contributions in 
$\qqb\to \ttb g$ and $\qqb\to \ttb$ each have an infrared (but no
collinear) singularity at ${\cal O}(\alpha_s^3)$~\cite{Kuhn}, which 
cancels in their sum 
and leaves behind a single logarithm, represented by the term
proportional to $1/(1-\th)_+$ in Eq.~(\ref{plus}). We will see in
the next section, however, that, starting with
the next order, leading logarithms contribute to the asymmetric cross section
as an overall factor.

\section{NLL Resummation}

The resummation of the soft-gluon contributions is organized in 
Mellin-$N$ moment space. We take a moment of the hadronic cross section
with respect to the variable $\tau$:
\begin{equation} \label{mellin}
\frac{d \sigma^N}{d\cos\theta} \equiv \int_0^1 d\tau \tau^{N-1}
\frac{d^2\sigma}{d\tau d\cos\theta} \; .
\end{equation}
Under Mellin moments, the convolutions in the factorized cross section  
near threshold become products:
\begin{equation} \label{nlogen}
\frac{d \sigma^N}{d\cos\theta}=\sum_{ab}f_a^{H_A,N}(\mu^2) 
f_b^{H_B,N}(\mu^2) \hat{\sigma}_{ab} (N,\theta) \; ,
\end{equation}
where the $f_a^{H_A,N},f_b^{H_B,N}$ are the Mellin moments of the 
parton distributions, defined by analogy to Eq.\ (\ref{mellin}),
 and where
\begin{eqnarray}
\hat{\sigma}_{ab} (N,\theta)&=& \int_0^1 d\hat{\tau}\th^{N-1}
\frac{d^2 \hat{\sigma}_{ab}(\th,\cos \theta)}
{d \hat{\tau} \,d \cos \theta} \; .
\end{eqnarray} 
The threshold limit  ($\th \to 1$) now corresponds to $N\to \infty$ 
in moment space. 

Threshold resummation results in exponentiation of the soft gluon 
corrections in moment space. 
Unlike color singlet cases like the Drell-Yan process, 
in heavy-flavor production soft gluons emitted at large angles interfere 
with the color structure of the underlying Born process. One must then 
take into account all color structures and sum over them. The details
of this procedure were worked out for scattering at fixed angles
in~\cite{KS,Kidonakis:1998ur,BCMN1}. For a given partonic
channel ($ab=q\bar{q}$, $gg$), the resumed perturbative cross section 
is given by \footnote{See, in particular, Eq.\ (50) of
Ref.\ \cite{Kidonakis:1998ur}.}
\beeq \label{resu}
\hat{\sigma}^{{\mathrm{(res)}}}_{a b} (N,\theta)
&=&C_{ab}(\theta) 
\,\Delta_{a}(N)  \Delta_{b}(N)\;  {\mathrm{Tr}} \left\{ H^{(0)}_{a b} 
(\theta)  \, 
\left[{\cal S}_{a b} (N,\theta) \right]^{\dagger}  \, S^{(0)}_{a b} \,  
{\cal S}_{a b} (N,\theta) \right\}  \; ,
\eeeq
where we have suppressed for simplicity the dependence of the 
various functions on the pair mass $M_{\ttb}$, but have kept
dependence on the scattering angle $\theta$ wherever it occurs.
Even before defining the various factors in the resummed cross
section, we may note that it naturally gives rise to a charge asymmetry.
In Eq.~(\ref{resu}), all dependence on the c.m. scattering
angle $\theta$ resides in the color trace part and in the coefficients
$C_{ab}$. Therefore, these generate the charge-asymmetric part
of the cross section near threshold:
\begin{eqnarray} \label{resuD}
\Delta \hat{\sigma}^{{\mathrm{(res)}}}_{q \bar{q}} (N,\theta)
&=&\,\left(\Delta_{q}(N)\right)^2\left[ C_{q\bar{q}}(\theta) 
\,  {\mathrm{Tr}} \left\{ H^{(0)}_{a b} 
(\theta)  \, 
\left[{\cal S}_{a b} (N,\theta) \right]^{\dagger}  \, S^{(0)}_{a b} \,  
{\cal S}_{a b} (N,\theta) \right\} \right. \nn \\
&&\left. -
C_{q\bar{q}}(-\theta) 
\,  {\mathrm{Tr}} \left\{ H^{(0)}_{a b} 
(\theta)  \, 
\left[{\cal S}_{a b} (N,-\theta) \right]^{\dagger}  \, S^{(0)}_{a b} \,  
{\cal S}_{a b} (N,-\theta) \right\}\right]
 \; .
\end{eqnarray}
It is the ratio of Eq.\ (\ref{resuD}) and its charge-averaged counterpart 
that defines the asymmetry.  We anticipate that, when expanding
the resummed expression to next-to-leading order, leading logarithms
cancel and the $C_2$ term in Eq.\ (\ref{plus}) is reproduced.

Now let us review the elements of the resummed cross sections.
The trace in Eq.~(\ref{resu}) is taken in a space of color exchange operators
\cite{KS,Kidonakis:1998ur}. 
At lowest order, $H^{(0)}_{a b}$ and $S^{(0)}_{a  b}$ are the 
hard-scattering and the zeroth-order soft functions, respectively. 
The factors ${\cal S}_{a b}$ are also matrices in color space 
and depend on the basis of color tensors used to describe color exchange.
Employing the $s$-channel singlet-octet basis
of~\cite{KS}, one has 
for the $q\bar{q}$ subprocess, which we are mostly interested in here,
\beqa
H^{(0)}_{q\bar q} &=& \as^2 \,  \left(
                \begin{array}{cc}
                 0 &0  \\[2mm]
                 0 &  2 \left( \frac{t^2+u^2}{s^2} + 
2 \frac{m_t^2}{s}\right)/C_A^2  \end{array} \right) \, ,
\nonumber \\[4mm]
S^{(0)}_{q\bar q} &=&  \left(
                \begin{array}{cc}
                 C_A^2 &0  \\[2mm]
                 0 &  (C_A^2-1)/4   \end{array} \right) \, ,
\eeqa
where $t\equiv (p_t-p_a)^2-m_t^2=-s (1-\beta\cos\theta)/2$, 
$u\equiv (p_{\bar{t}}-p_a)^2-m_t^2=-s (1+\beta\cos\theta)/2$, 
with $p_a$ the momentum of initial parton $a$, again 
$\beta=\sqrt{1-4 m_t^2/s}$,
and $C_A=3$. The corresponding expressions for the soft anomalous 
dimension matrices of the $gg$-initiated subprocess may
be found in~\cite{KS}. Note that the Born cross sections are 
recovered by computing ${\mathrm{Tr}}\{H^{(0)}_{ab} S^{(0)}_{ab}\}$ and
that $H^{(0)}_{ab}$ is symmetric under interchange of $t$ and $u$ and
hence charge symmetric.

Each of the functions 
$\Delta_{a,b}(N)$ and ${\cal S}_{a b} (N,\theta)$ is an exponential. 
$\Delta_a(N)$ represents the effects of soft-gluon 
radiation collinear to initial parton $a$ and
in the $\overline{\rm{MS}}$ scheme is given by 
\begin{align}\label{Dfct}
\ln \Delta_a(N)&=  \int_0^1 \f{z^{N-1}-1}{1-z} 
\int_{\mu^2}^{(1-z)^2 M_{t\bar{t}}^2} \f{dq^2}{q^2} A_a(\as(q^2)) \; ,
\end{align}
and similarly for $\Delta_b(N)$. 
The function
$A_a$ is a perturbative series in $\as$,
\begin{equation}
A_a(\as)=\frac{\as}{\pi} A_a^{(1)} +  
\left( \frac{\as}{\pi}\right)^2 A_a^{(2)} + \ldots \; ,
\end{equation}
with~\cite{KT}:
\begin{equation} 
\label{A12coef} 
A_a^{(1)}= C_a
\;,\;\;\;\; A_a^{(2)}=\frac{1}{2} \; C_a  \left[ 
C_A \left( \frac{67}{18} - \frac{\pi^2}{6} \right)  
- \frac{5}{9} N_f \right]\; ,
\end{equation}  
where $N_f$ is the number of flavors, and 
\begin{eqnarray} 
&&C_q=C_F=(N_c^2-1)/2N_c=4/3  \;, \;\;\;C_g=C_A=N_c=3 \; .
\end{eqnarray} 
The factors $\Delta_{a,b}(N)$ generate leading threshold enhancements,
due to soft-collinear radiation, as the same overall factors in 
the charge symmetric and  antisymmetric cross sections.

The large-angle soft gluon exponentials ${\cal S}_{a b} (N,\theta)$ are 
dependent on the process and mix the color structure. One has
\beeq
\label{GammaSoft}
{\cal S}_{a b} \left(N,\theta
\right) &=& {\cal P}\exp\left[ 
\frac{1}{2} \int^{M_{t\bar{t}}^2/N^2}_{M_{t\bar{t}}^2} 
\frac{d q^2}{q^2} \Gamma^S_{a b} 
\left(\as(q^2),\theta \right)\right] \, ,
\eeeq
where ${\cal P}$ denotes path ordering and where
$\Gamma^S_{a b}$ are soft anomalous dimensions, which are 
also matrices in a given color basis. They are perturbative; for the
resummation at NLL one only needs the first-order term and path
ordering becomes irrelevant. For the $q\bar{q}$ subprocess, the first-order
anomalous dimension matrix can be represented, in the $s$-channel 
singlet-octet basis, as
\begin{equation}
\Gamma_{q\bar{q}}^{S,(1)}=\frac{\as}{\pi}\left(\begin{array}{cc}
\Gamma^{q\bar{q}}_{11} & \Gamma^{q\bar{q}}_{12} \\
\Gamma^{q\bar{q}}_{21} & \Gamma^{q\bar{q}}_{22}
\end{array}
\right) \, ,
\label{matrixqqQQ}
\end{equation}
with matrix elements~\cite{KS}\footnote{Note that for our definition
of the charge asymmetry we need to interchange $t$ and $u$ in the
results of~\cite{KS}.}
\begin{eqnarray}
\Gamma^{q\bar{q}}_{11}&=&-C_F \, [L_{\beta}
+1+i\pi] \, ,
\nonumber \\
\Gamma^{q\bar{q}}_{21}&=&2 \ln\left(\frac{t}{u}\right) \, ,
\nonumber \\
\Gamma^{q\bar{q}}_{12}&=&
\frac{C_F}{C_A} \ln\left(\frac{t}{u}\right) \, ,
\nonumber \\
\Gamma^{q\bar{q}}_{22}&=&C_F
\left[4\ln\left(\frac{t}{u}\right)
-L_{\beta}-1-i\pi\right]
\nonumber \\ &&
{}+\frac{C_A}{2}\left[-3\ln\left(\frac{t}{u}\right)
-\ln\left(\frac{m_t^2 s}{tu}\right)+L_{\beta}+1+i\pi \right]\, ,
\label{GammaqqQQ}
\end{eqnarray}
where
\begin{equation}
L_{\beta}=\frac{1-2m_t^2/s}{\beta}\left(\ln\frac{1-\beta}{1+\beta}
+i\pi  \right)\, .
\end{equation}
Finally, the coefficients $C_{ab}(\theta)$ contain $N$-independent hard 
contributions arising from one-loop virtual corrections. They are 
perturbative as well, and have the expansion 
\begin{eqnarray} 
C_{ab}(\theta) = 1 + \frac{\as}{\pi} C_{ab}^{(1)}(\theta) 
+ {\cal O}(\as^2) \; .
\end{eqnarray} 
The coefficients $C_{ab}^{(1)}$ may be obtained by comparison of the
resummed formula to the full NLO calculation. We note that they contain 
the Coulomb corrections
which diverge as $1/\beta$ at $s \sim 4 m_t^2$.
As indicated, the $C_{ab}^{(1)}$ depend on the scattering angle 
$\theta$, and in fact for the $q\bar{q}$ subprocess they also contain 
a charge-asymmetric part. The full coefficients have been derived in 
Ref.~\cite{KLMV} and are given by very lengthy expressions. 
Starting from slightly corrected versions \footnote{Specifically, we
use Eq.\ (A.19) of Ref.\ \cite{FMNR} in the
second and
third integral in (A.20) of Ref.\ \cite{MNR}
and in Eqs.\ (A.9) and (A.11) of the first paper of Ref.\ \cite{KLMV}.} of the
expressions given in~\cite{KLMV}, we have been able 
to verify that the charge-asymmetric part of the resulting coefficient 
for the $q\bar{q}$ process reproduces the corresponding result
given in~\cite{Kuhn}.

We now give explicit formulas for the expansions of the 
resummed exponents to NLL accuracy following the general approach 
of \cite{Catani:1996yz}. The functions $\Delta_{a,b}(N)$ become
\beeq
\label{lndeltams}
\!\!\! \!\!\! \!\!\! \!\!\! \!\!\!
\ln \Delta_a(N,\as(\mu^2),M_{t\bar{t}}^2/\mu^2) 
&\!\!=\!\!& \ln \bar{N} \;h_a^{(1)}(\lambda) +
h_a^{(2)}(\lambda,M_{t\bar{t}}^2/\mu^2) + 
{\cal O}\left(\as(\as \ln N)^k\right) \,.
\eeeq
Here $\lambda=\b0 \as(\mu^2) \ln \bar{N}$ with $\bar{N}=
N {\mathrm{e}}^{\gamma_E}$, where $\gamma_E$ is the Euler constant. 
The functions $h^{(1,2)}$ are given by 
\begin{align} 
\label{h1fun}
h_a^{(1)}(\la) =& \f{A_a^{(1)}}{2\pi \b0 \la} 
\left[ 2 \la+(1-2 \la)\ln(1-2\la)\right] \;,\\ 
\label{h2fun}
h_a^{(2)}(\la,M_{t\bar{t}}^2/\mu^2) 
=&-\f{A_a^{(2)}}{2\pi^2 \b0^2 } \left[ 2 \la+\ln(1-2\la)\right]
+ \f{A_a^{(1)}}{2\pi \b0}\ln(1-2\la)
\ln\f{M_{t\bar{t}}^2}{\mu^2} \nn \\ 
&+ \f{A_a^{(1)} \bone}{2\pi \b0^3} 
\left[2 \la+\ln(1-2\la)+\f{1}{2} \ln^2(1-2\la)\right]\;,  
\end{align} 
where $\b0 = \left( 11 C_A - 2 N_f \right)/12\pi$,
and 
\begin{align}
\bone=  \frac{1}{24 \pi^2} 
\left( 17 C_A^2 - 5 C_A N_f - 3 C_F N_f \right) \;\; .
\label{bcoef}
\end{align}
The function $h_a^{(1)}$ above contains all LL terms in the perturbative 
series, while $h_a^{(2)}$ is of NLL only. For
a complete NLL resummation one also needs the expansion of the integral
in Eq.~(\ref{GammaSoft}), given by,
\beeq
\label{GammaSoft1}
\ln {\cal S}_{a b} \left(N,\theta
\right) &=& \frac{\ln(1 - 2 \lambda)}{2\pi b_0} \;
\Gamma_{q\bar{q}}^{S,(1)} (\theta)\, .
\eeeq

While the full structure of the resummed expressions is rather long and 
complicated, a major simplification occurs when one expands it for 
small $\beta$ and ignores the coefficient $C_{ab}$. One finds in this case
\begin{equation} \label{resuD1}
\hat{\sigma}^{{\mathrm{(res)}}}_{q \bar{q}} (N,\theta)
=\,\hat{\sigma}_{q \bar{q}}^{{\mathrm{(Born)}}}(\theta)\,
\left(\Delta_{q}(N)\right)^2 
\left\{ 1+
\frac{\beta\cos\theta (8 C_F-3 C_A) \ln (1-2 \lambda)}{\pi b_0}\right\}\,
{\mathrm{e}}^{-\frac{C_A}{2 \pi b_0}\ln (1-2 \lambda)}  \; .
\end{equation}
Here, the factor $8 C_F-3 C_A$ is the typical color factor
associated with the QCD charge asymmetry~\cite{Kuhn}. 
One can see how the single threshold logarithm arises at the first 
order in $\as$ in the charge-asymmetric part. The charge-asymmetric piece
is suppressed by the factor $\beta$, but enhanced by the term
$\ln (1-2 \lambda)$. All factors outside
the curly brackets are common to the charge-asymmetric and the
charge-summed parts and are expected to largely cancel in the
charge asymmetry at hadron level. We note that in the limit $\beta\to 0$ 
our formulas above reproduce the moment-space expressions for the 
resummed total heavy-flavor cross section derived in~\cite{BCMN}.

In our discussion below, we use the full formula~(\ref{resuD}) when 
calculating the charge 
asymmetry. Since the matrices involved for the $q\bar{q}$ subprocess
are two-dimensional, it is straightforward to perform the required
exponentiations and other manipulations,
explicitly employing a diagonal color basis for these $2\times 2$ 
matrices \cite{KS}. For the (charge-symmetric)
$gg$ subprocess, the matrices are three-dimensional, and this 
procedure becomes more complicated. We found it simpler here to 
do the matrix manipulations numerically, calculating in particular
the matrix exponentials by expanding the exponential series to the
tenth order. We emphasize again that the $gg$ process is 
charge-symmetric. Thus, it only contributes to the denominator
of the charge asymmetry, diluting the asymmetry 
somewhat, because the effects of threshold resummation
can be larger for gluons than for quarks.

\section{Phenomenological Results}

We will now investigate the numerical size of the QCD charge asymmetry
for top quark production at the Tevatron, making use of the 
resummation formulas presented above. In order to do this, we first 
need to specify the inverse Mellin transform. This
requires a prescription for dealing with the singularity
in the perturbative strong coupling constant in the resummed
exponent. We will use the {\em Minimal Prescription} developed 
in Ref.~\cite{Catani:1996yz}, which relies on use of the NLL expanded 
forms Eqs.~(\ref{lndeltams})-(\ref{h2fun}), and on choosing
a Mellin contour in complex-$N$ space that lies to the {\it left}
of the poles at $\lambda=1/2$ in the Mellin integrand:
\begin{align}
\label{hadnmin}
\frac{d^2 \sigma}{dM^2_{t \bar{t}}  \, d \cos \theta}=
\;\int_{C_{MP}-i\infty}^{C_{MP}+i\infty}
\;\frac{dN}{2\pi i} \;\tau^{-N}
\sigma^{\rm (res)}(N) \; ,
\end{align}
where $b_0\as(\mu^2)\ln C_{MP}<1/2$, but all other poles
in the integrand are as usual to the left of the contour. The
result defined by the minimal prescription has the property that 
its perturbative expansion is an asymptotic series that 
has no factorial divergence and therefore
no ``built-in'' power-like ambiguities. Power corrections may
then be added, as phenomenologically required.

For our calculations we use the CTEQ6M parton
distributions~\cite{cteq}. To obtain these in Mellin-moment space,
we follow~\cite{ddfz} and perform a simple fit to each parton distribution 
at each scale needed, using a functional form that allows Mellin
moments to be taken analytically. We note that for the Tevatron case 
considered here, $t\bar{t}$ pairs are largely produced in valence-valence
scattering, for which the parton distributions are rather well known.
The higher-order and resummation effects will be very similar for
other sets of parton distributions. We use $m_t=170.9$~GeV~\cite{cdfd0}, 
$\sqrt{S}=1.96$~TeV, and our default choice for the 
factorization/renormalization scale is $\mu=M_{t\bar{t}}$. 

\begin{figure}[htb]
\begin{center}
\vspace*{0.6cm}
\epsfig{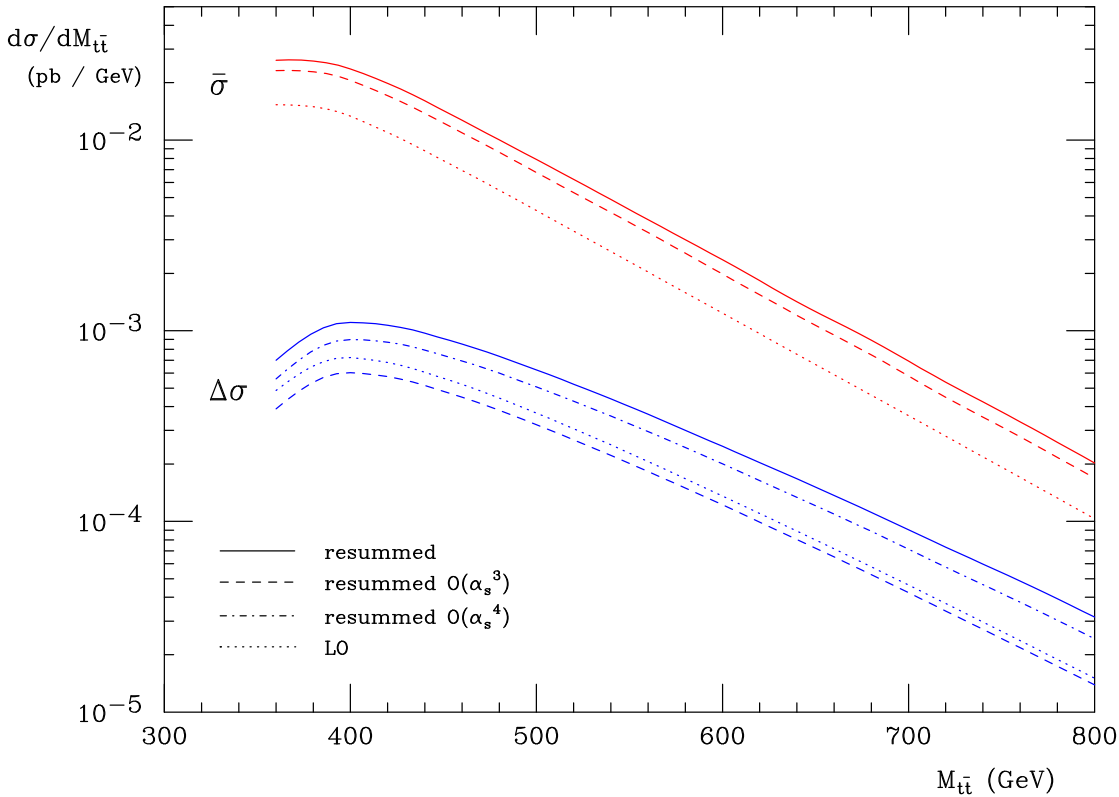}
\end{center}
\vspace*{-.5cm}
\caption{Charge asymmetric and charge averaged cross sections 
$d \Delta \sigma/dM_{t \bar{t}}$ and $d \bar{\sigma}/dM_{t \bar{t}}$
as functions of the $t\bar{t}$ pair mass, integrated over 
$0\leq\cos\theta \leq 1$. Dotted lines are LO, dashed lines include
the first-order corrections generated by resummation, and solid lines
show the full resummed result. For the charge-asymmetric part, 
we also show the second-order expansion of the resummed cross section,
which is of ${\cal O}(\as^4)$, because this contains the first
non-trivial QCD correction in this case. 
\label{fig1} }
\vspace*{0.cm}
\end{figure}

The lower set of lines in Fig.~\ref{fig1} shows our results for 
the charge asymmetric cross section $d \Delta \sigma/
dM_{t \bar{t}}$ as a function of the $t\bar{t}$ pair mass, while 
the upper set presents the charge-averaged one, $d \bar{\sigma}/
dM_{t \bar{t}}$. We show here the cross sections integrated over 
$0\leq\cos\theta \leq 1$. In both cases, we show by the 
{\it dotted lines} the lowest-order result, which is ${\cal O}(\as^2)$ in the 
charge-averaged case, and ${\cal O}(\as^3)$ in the asymmetric part.
In the former case, this result is just based on the usual
Born cross processes, $q\bar{q}\to t\bar{t}$ and $gg\to t\bar{t}$.
For the charge asymmetric part, we make use of the expressions given 
in the Appendix of~\cite{Kuhn}. This includes the small contribution by
the quark-gluon flavor excitation process $qg\to t\bar{t}q$.
Next in Fig.~\ref{fig1} we show the first-order expansion  of the resummed 
cross sections ({\it dashed lines}), which are approximations to the full 
NLO result. 
In case of the charge-asymmetric piece, which only starts at NLO, we 
can check the quality of this approximation by comparing the dotted and 
dashed lines. One can see that the two results agree very well
in this case,
implying that the threshold corrections addressed by resummation dominate. 
\footnote{In principle, one may carry out the same check for the
charge-symmetric cross section at fixed $M_{t\bar{t}}$
and $\theta$ by comparing to the full NLO calculations
of \cite{Beenakker:1988bq,Nason}.   Such a study is unlikely
to change our results here qualitatively, and would go beyond
the scope of this work.}

For the charge-asymmetric part, we also show 
({\it dashed-dotted lines}) the second-order expansion of the resummed 
cross section,
which is of ${\cal O}(\as^4)$, because this contains the first
non-trivial QCD correction in this case. 
The solid lines finally show the full NLL-resummed result. One can
see that resummation has a very significant impact on the predicted
cross sections, in particular for the charge-asymmetric part at
high  $M_{t \bar{t}}$. Note that for the resummed curve for this
part we have performed a matching to the full ${\cal O}(\as^3)$ 
result of~\cite{Kuhn} by correcting it by the difference between the
dotted and dashed lines. In this way, the ${\cal O}(\as^3)$ is taken 
into account in full, and the soft-gluon contributions beyond NLO are 
resummed to NLL.

\begin{figure}[htb]
\begin{center}
\vspace*{-0.6cm}
\epsfig{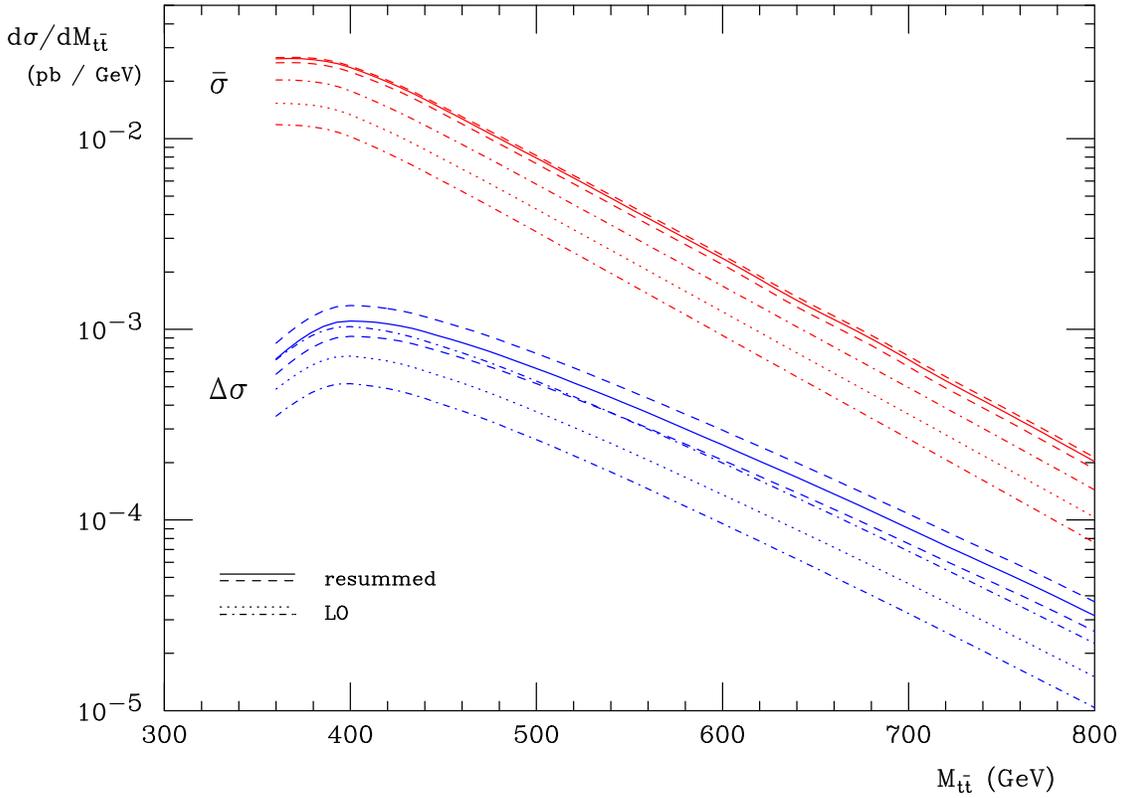}
\end{center}
\vspace*{-.5cm}
\caption{Scale dependence of the LO (dotted and dash-dotted)
and NLL resummed (solid and dashed) cross sections, for a scale 
variation $M_{t \bar{t}}/2\leq\mu\leq 2 M_{t \bar{t}}$. \label{fig2} }
\vspace*{0.cm}
\end{figure}

In Fig.~\ref{fig2} we analyze the scale dependence of the results,
for both the charge asymmetric and averaged parts. The dash-dotted
lines correspond to a variation of $M_{t \bar{t}}/2\leq\mu\leq
2 M_{t \bar{t}}$, with the central lines (dotted) the ones for $ \mu=
M_{t \bar{t}}$ already shown in Fig.~\ref{fig1}. The higher results
are obtained for the lower choice of scale. The dashed and solid
lines display the same for the resummed cross sections. One can 
see a very significant reduction of scale dependence, in particular
for the charge averaged cross section. This improvement in scale
dependence due to threshold resummation is in line with similar findings in 
the literature for other cross sections~\cite{BCMN,KLMV,Matteo,scale1,scale}. 
We note that the fact that $t\bar{t}$ production at the Tevatron proceeds 
primarily through $q\bar{q}$ valence-valence annihilation helps here, since 
only the flavor-diagonal non-singlet evolution matters, which is
part of the resummation formula~\cite{BCMN,scale1}.

\begin{figure}[htb]
\begin{center}
\vspace*{-0.6cm}
\epsfig{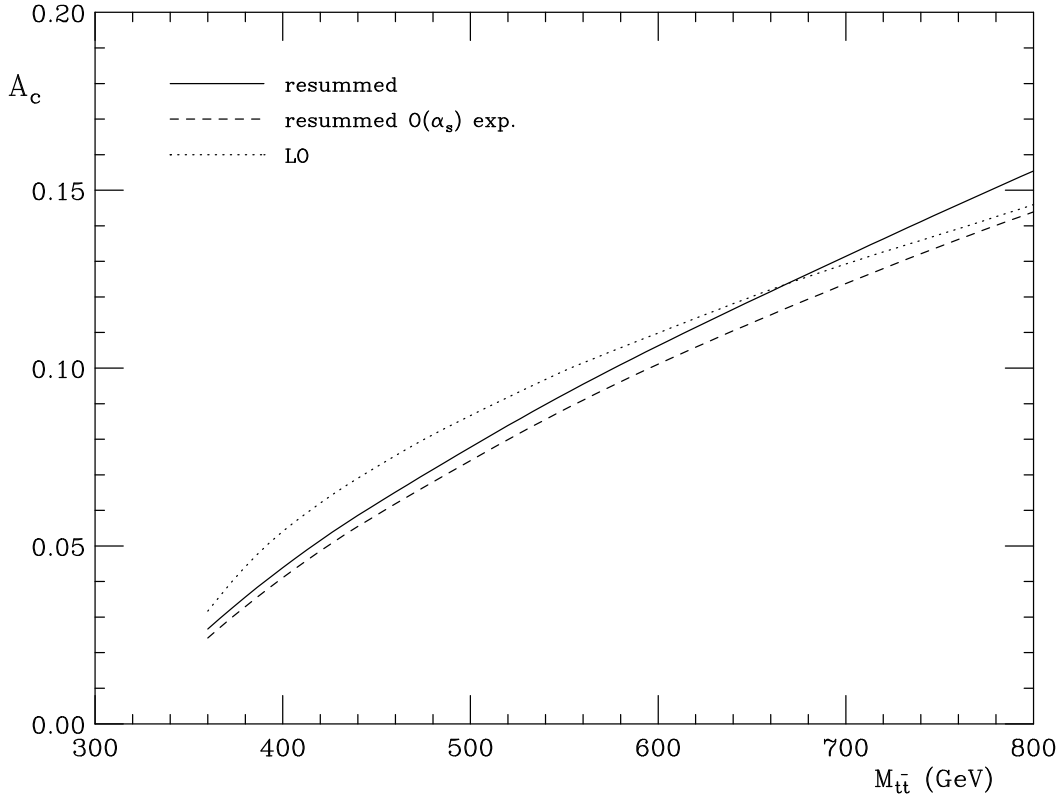}
\end{center}
\vspace*{-.5cm}
\caption{Charge asymmetry corresponding to the curves in Fig.~\ref{fig1}.
\label{fig3} }
\vspace*{0.cm}
\end{figure}

Figure~\ref{fig3} shows the charge asymmetries $A_c=d\Delta \sigma/
d\bar{\sigma}$ corresponding to the various curves in Fig.~\ref{fig1},
as functions of the pair mass. 
As before, the dotted line shows the LO result, the solid represents the 
full resummed result, and the dashed one is the expansion of the resummed 
cross section. For the latter we expand the cross section to 
${\cal O}(\as^4)$ in the numerator of the asymmetry, and to 
${\cal O}(\as^3)$ in the denominator, thus taking into account the
first non-trivial QCD correction in both cases. Had we expanded
both numerator and denominator to ${\cal O}(\as^3)$, the numerator
would be at LO, and an artificially small asymmetry would result. 
One can see that the various results are rather
close, implying that the net effect of resummation on the charge
asymmetry is not large. This is related to the fact that
the double-logarithmic factors $\propto (\Delta_q(N))^2$ 
in the resummation formula are
the same for the charge asymmetric and averaged parts. Towards
lower $M_{t \bar{t}}$, where the speed of the produced top quark
becomes small, all asymmetries become small on account of 
Eq.~(\ref{resuD1}). At large pair masses, the additional single threshold
logarithm enhances the asymmetry. 

\begin{figure}[htb]
\begin{center}
\vspace*{-0.6cm}
\epsfig{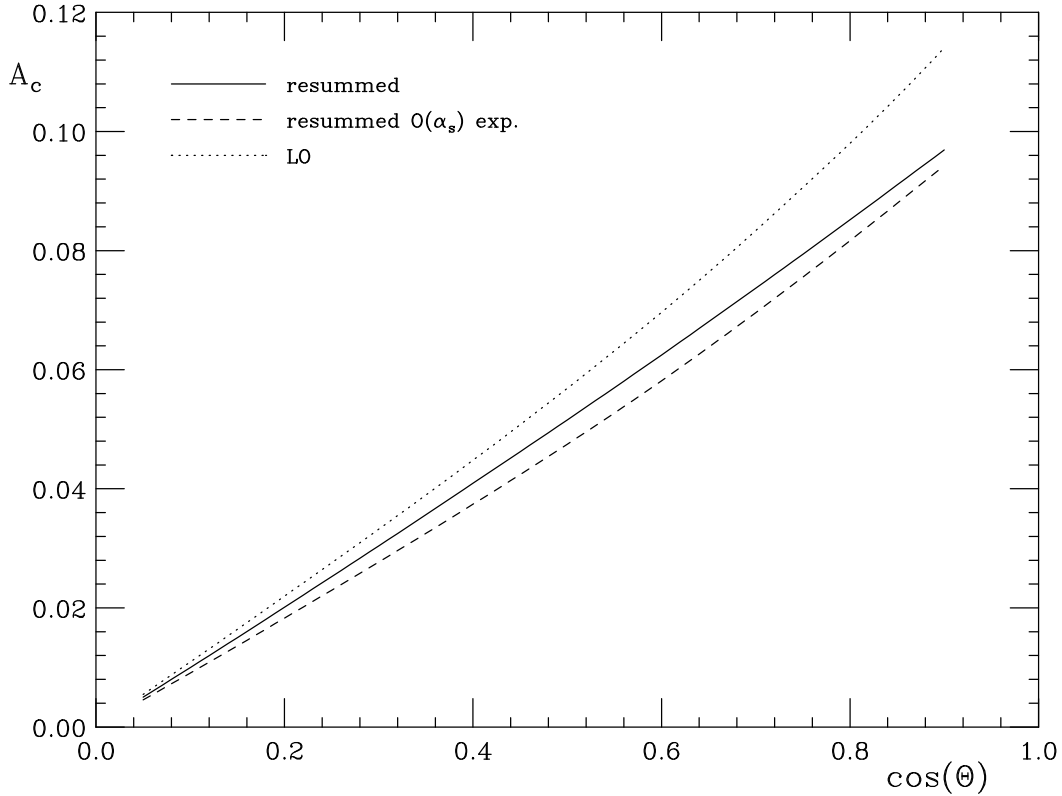}
\end{center}
\vspace*{-.5cm}
\caption{Same as Fig.~\ref{fig3}, but as a function of $\cos\theta$,
integrated over the $t\bar{t}$ pair mass. \label{fig4} }
\vspace*{0.cm}
\end{figure}

In Fig.~\ref{fig4} we consider the asymmetry as a function of 
$\cos\theta$, with $M_{t \bar{t}}$ integrated over the allowed
kinematic region. Again the net effect of resummation on the 
asymmetry is relatively moderate. Integration over $M_{t \bar{t}}$ 
leads generally to a smaller
asymmetry, because lower pair masses, at which the asymmetry
decreases when going from the LO to the resummed case (see Fig.~\ref{fig3}), 
dominate the cross section. 
As Figs.\ \ref{fig3} and \ref{fig4} show, the resummed asymmetries 
grow substantially with both pair mass, when integrated over rapidity,
and with the relative rapidity of the pair.
Interestingly, these results are consistent with 
the explicit NLO results presented in Refs.\ \cite{Bowen,Dittmaier},
which indicate a decrease in the charge asymmetry,
and even a reversal of its sign, for top pair plus jet
cross sections.    In such final states, the NLO virtual
corrections to inclusive pair production are absent,
and it is the latter corrections that determine the sign
of the asymmetry itself.   

We finally turn to the total charge asymmetry $A_c^{\mathrm{tot.}}$, 
integrated over $M_{t \bar{t}}$ and $0\leq\cos\theta \leq 1$. 
At LO, using the scale $\mu=M_{t\bar{t}}/2$, 
we find $A_c^{\mathrm{tot.}}=6.7\%$. Resummation results in only 
a small change, $A_c^{\mathrm{tot.}}=6.6\%$. We note that when
varying the scale over the range $m_t\leq \mu\leq M_{t\bar{t}}$,
the LO charge-symmetric part of the cross section varies by about $\pm 20\%$
around its central value, which is improved by resummation to a variation
of about $\pm 3\%$. The scale dependence of the asymmetric part of the cross 
section improves from $\pm 28\%$ to $\pm 13\%$. The resummed asymmetry shows
a variation over this range of scales of about $\pm 12\%$. Thus our results 
for the higher-order corrections to $A_c^{\mathrm{tot.}}$ are
well consistent with the estimate of a $\sim 30\%$ uncertainty
made in Ref.\ \cite{Antunano:2007da}.

\section{Conclusions and Outlook}

We have presented a study of the next-to-leading logarithmic 
QCD threshold resummation effects on the charge asymmetry in inclusive
$t\bar{t}$ production at the Tevatron. We have found that 
the asymmetry is stable with respect to the higher-order corrections
generated by threshold resummation. We have also found that
resummation significantly decreases the dependence of the results
on the factorization and renormalization scales, thus making the
Standard Model prediction for the asymmetry more reliable. 

It will be interesting to extend these studies to the case
of $t\bar{t}\,{\mathrm{jet}}$ production, for which sizable
negative NLO corrections have been found~\cite{Dittmaier}.
Also, there will be interesting applications at the LHC in
situations near partonic threshold, i.e., when the $\ttb$ pair mass
becomes of the order of 1~TeV or larger. Because the initial $pp$ 
state is symmetric, one needs to apply additional cuts
(for example, on the $\ttb$ pair rapidity) here in order to generate 
a non-vanishing charge asymmetry~\cite{Kuhn}.

\section*{Acknowledgments} 

We are grateful to E.\ Laenen for very helpful communications
and to D.\ de Florian, G.\ Perez, G.\ Rodrigo,
J.\ Smith, and M.\ Stratmann for useful discussions. 
W.V.\ is grateful to the U.S.\ Department of Energy 
(contract number DE-AC02-98CH10886) for
providing the facilities essential for the completion of his work.
This work was supported in part by the National Science Foundation, 
grants PHY-0354776, PHY-0354822 and PHY-0653342.


\end{document}